# Fabrication of micro-structures for optically driven micromachines using two-photon photopolymerization of UV curing resins

**Theodor Asavei, Timo A. Nieminen, Norman R. Heckenberg and Halina Rubinsztein-Dunlop**

Centre for Biophotonics and Laser Science, School of Physical Sciences, The University of Queensland, Brisbane QLD 4072, Australia

E-mail: asavei@physics.uq.edu.au

**Abstract.** Two-photon photopolymerization of UV curing resins is an attractive method for the fabrication of microscopic transparent objects with size in the tens of micrometers range. We have been using this method to produce three-dimensional structures for optical micromanipulation, in an optical system based on a femtosecond laser. By carefully adjusting the laser power and the exposure time we were able to create micro-objects with well-defined 3D features and with resolution below the diffraction limit of light. We discuss the performance and capabilities of a microfabrication system, with some examples of its products.

## 1. Introduction

The first fabricated microstructures appeared at the end of the 1980s and the beginning of the 1990s. They were the microelectromechanical systems (MEMS) commonly known as micromachines. MEMS are electromechanical machines that range in size from a micrometer to a millimetre. They are built using techniques based on integrated circuit (IC) fabrication methods.

The first reported MEMS were silicon electrical micromotors with a diameter of 100µm [1, 2]. Nowadays MEMS are used as sensors and actuators in various applications. There are several methods of producing MEMS. The oldest one is bulk micromachining [3], and it is used to fabricate silicon based MEMS. This method creates micro-mechanical structures by etching into a silicon wafer, using the whole thickness of the wafer.

As opposed to bulk micromachining, surface micromachining [3] was created in the late 1980s in order to obtain planar MEMS which could be compatible with on-chip integrated circuits. This technique creates MEMS elements from deposited thin films on the surface of a silicon wafer. Polycrystalline silicon, silicon nitride or silicon dioxide films can be deposited layer-by-layer on the silicon wafer, photo-patterned with photolithography and selectively removed through etching to obtain the desired machine elements.

However, a drawback of surface micromachining is the fact that it is limited to the production of two-dimensional structures. In other words one cannot create three-dimensional structures with high aspect ratio, or curved surfaces with surface micromachining. Fortunately, apart from the above mentioned methods, micromachines can be fabricated optically by means of microstereolithography. This method emerged as a consequence of the rapid prototyping techniques which became available in the 1980s, namely the stereolithography technique [4] and it has the advantage that it can create real 3D microstructures. It is based on the fact that certain resins can be hardened by exposure to UV light. Basically the 3D microstructure is fabricated by stacked 2D hardened elements, which were initially obtained from slicing the 3D shape of the structure in a computer-aided design (CAD) program.

Other advantages of this method are the biocompatibility of the hardened resin and the fact that the produced objects can be easily tailored in size. Apart from being optically fabricated, these microstructures can be optically driven. Optical drive is based on the fact that microscopic particles (25 nm to 10 µm) can be stably trapped in a tightly focused laser beam due to gradient forces [5]. The advantage of optical drive is that no contact is required so that it is particularly well suited to biological applications. Forces on the order of pN with mW of laser power can be exerted on the trapped particles as well as torques on the order of pNµm through exchange of optical angular momentum between the trapped object and the trapping beam. Such optical angular momentum is carried in 'spin' (circular polarization) or 'orbital' (helical phase structure) forms.

The first optically fabricated microstructures reported in 1993 were a micro valve, a micro coil spring with a 250 µm height and 50 µm coil diameter and a micro pipe with 30 µm inner diameter [6]. They were fabricated with a Xenon lamp as UV source focused in a liquid resin. The resin hardened only in the focal spot of the UV beam and by scanning the sample over the focus 3D objects could be obtained.

The resolution of the optically fabricated microstructures is determined by the size of the smallest solidified volume, called "voxel" (**vo**lumetric pi**xel**). For the above mentioned objects the resolution was 5X5X3 µm$^3$. The resolution can be substantially increased if instead of using one-photon absorption of UV light one uses two-photon absorption of IR light. The increase in resolution is due to the fact that two-photon absorption probability is proportional to the square of light intensity and hence the resin polymerizes in a far smaller volume than the one in one-photon absorption.

The two-photon photopolymerization (2PP) technique was pioneered by J. Strickler and W. Webb in 1991, following the application of two-photon excitation in two-photon laser scanning fluorescence microscopy [7]. Two-photon excitation was used to record high-density digital information in a multilayered 3D format. The information was written in submicrometre size voxels in a photopolymer with femtosecond IR laser pulses [8].

The first 3D microfabricated structures made by two-photon photopolymerization were reported in 1997 [9]. They were spiral structures with a diameter of 6 µm and a coil width of 1.3 µm.

Since then, various micromachines have been produced (micropumps, microgears, microneedles) with high resolution [10-12]. A lot of research is being done in order to improve the spatial resolution of the photopolymerization process and 3D photonic crystal structures with sub-100 nm resolution have been reported [13,14].

Complex 3D objects of potentially any shape can be fabricated by this method for optical trapping experiments. However, sub-wavelength resolution is already adequate to allow us to tailor the shape of the microstructure in order to optimize the transfer of linear and/or angular momentum between the trapping beam and the trapped object. One can also produce micrometer sized devices which can act as micro-holograms and change the properties of an incident laser beam creating, for example, a beam carrying orbital angular momentum in the form of a Laguerre-Gauss beam.

Another example is the combination of optical micromanipulation and 2PP for non-contact microassembly [15]. In this case 2PP can be used as a tool to build complementary planar microstructures which can be assembled by multi-beam optical traps.

## 2. Two-photon photopolymerization

The fact that some resins harden under UV exposure is based on photopolymerization of monomers in the liquid resin. The photopolymerization process is a particular type of radical chain polymerization.

Radical chain polymerization is a polymerization mechanism in which an initiator molecule produces a reactive centre in the form of a free radical. Polymerization occurs as a chain reaction by successive addition of monomers to the reactive centre. The chain reaction consists of three steps: initiation, propagation and termination [16].

The initiation step consists of two reactions. The first reaction is the homolytic dissociation of the initiator molecule I to obtain a pair of free radicals R+. In the case of photopolymerization, the dissociation is produced by absorption of light photons:

$$I + h\nu \rightarrow I^* \rightarrow 2R^*$$

The second reaction of initiation is the addition of the radical to the first monomer molecule M producing the chain-initiating radical $M_1+$:

$$R^* + M \rightarrow M_1^*$$

The propagation step consists of the successive additions of monomer molecules to $M_1^*$:

$$M_1^* + M \rightarrow M_2^* \ldots M_n^* + M \rightarrow M_{n+1}^*$$

Termination occurs when the radical centres are annihilated by combination (coupling) reaction between radicals:

$$M_n^* + M_m^* \rightarrow M_{n+m}$$

$M_{n+m}$ is the final product of the photopolymerization process which produces the hardened resin.

We use the NOA series of UV curing resins (Norland Products Inc., Cranbury, NJ, USA) to fabricate microstructures. These resins are based on a mixture of photoinitiator molecules and thiol-ene monomers and they are photopolymerized when exposed to light in the UV range with $\lambda$ <400 nm and require an energy flux of 2 to 4.5 J/cm$^2$ for a full cure.

In our experiments the two-photon polymerization is performed in an in-house inverted microscope. A scheme of the setup is shown in figure 1.

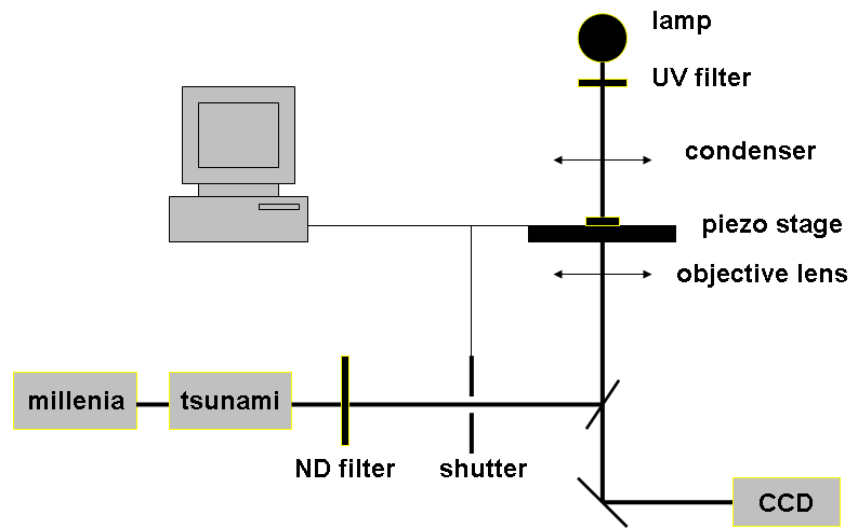

**Figure 1.** Two-photon photopolymerization experimental setup.

For two-photon polymerization we use the infrared light (λ=780 nm) produced by a femtosecond pulse Ti:Sapphire laser (Tsunami, Spectra Physics) pumped by a 532 nm solid state laser (Millenia, Spectra Physics). The pulse length is 80 fs with 80 MHz repetition rate.

The laser beam is attenuated to the power needed for polymerization and then passes through a computer controlled shutter and is reflected into the objective lens by a dichroic mirror. The objective lens is an Olympus 100X oil immersion lens with high numerical aperture (N.A. =1.3) to achieve high spatial resolution for polymerization.

The sample is mounted on a computer controlled piezo stage (model P-611.3S, Physik Instrumente) and is imaged onto a CCD with the same objective lens. The travel range of the piezo stage is 100 μm in all X, Y and Z directions.

The resin sample is sandwiched between two glass coverslips which are separated by an adhesive spacer with a thickness of 120 μm. 3D structures are fabricated by raster scanning the resin sample over the laser beam using the piezo stage. A scheme of the fabrication method is shown in figure 2.

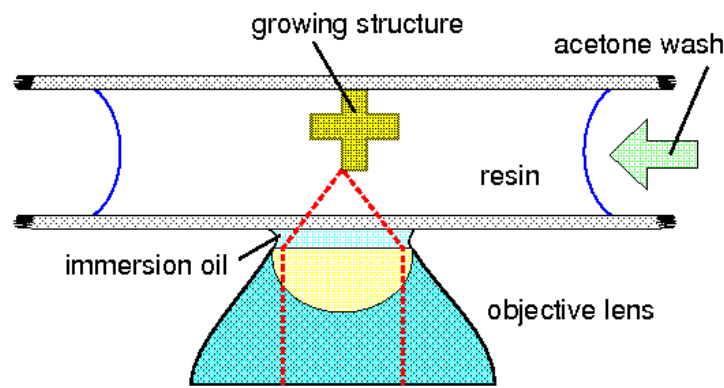

**Figure 2.** Schematics of the microfabrication method.

The 3D object is sliced into 2D layers (bitmap files) corresponding to the areas that need to be scanned. The program controlling the scanning stage reads the bitmap files and the resin is exposed (the shutter is opened) when the pixel in the bitmap is black (has value 0). 3D structures are obtained by moving the sample in the Z direction after each XY scan.

The bitmap resolution is set to 100X100 pixels which corresponds to 10X10 $\mu m^2$ travel in X and Y directions so that each individual pixel is 100X100 $nm^2$ in size. The offset step in the Z direction is 200 nm.

The structures are grown upside down on the upper cover slip. This top down scanning method has the advantage that the laser beam does not pass through already exposed resin.

After the polymerization, the unexposed resin is washed off with acetone, leaving the 3D structure attached to the cover slip.

The 3D structures are characterized with a scanning electron microscope (SEM) and bright field optical microscope.

**3. Spatial resolution**

The spatial resolution of the polymerization process is given by the size of a single voxel. Theoretically the voxel is a prolate spheroid and this follows from the calculation of the diffraction limited spot for the two-photon process which has this shape. This calculation was performed [17] and it shows that for a 1.3 N.A. objective lens the ratio between the long axis of the spheroid (which determines the axial resolution) and the short axis (the lateral resolution) is 3:1. Even though the shape of the single voxel sets a low axial resolution one can often effectively increase the axial resolution by overlapping layers in the Z direction.

Experimentally it is found that the resolution is dependent on the exposure time as well as the power of the laser beam.

However in order to be able to achieve the maximum resolution possible one needs to have a quantitative relation between the size of the voxel and the two parameters.

The length and diameter of the voxel can be calculated starting from the rate equation describing the production of radicals from the initiator molecules [18]:

$$\partial \rho / \partial t = (\rho_0 - \rho)\sigma_2 N^2 \qquad (1)$$

where $\rho$ is the density of radicals, $\sigma_2$ is the two-photon absorption cross-section, N is the photon flux and $\rho_0$ is the density of initiators.

By assuming a Gaussian distribution in the focal plane N becomes:

$$N = N_0 \exp(-2r^2 / r_0^2) \qquad (2)$$

The voxel diameter *d* can be calculated from (1) and (2) taking into account that the resin polymerizes in the region where the radical density is higher than the threshold radical density $\rho_{th}$:

$$d(P,t) = r_0 [\ln(CP^2 t)]^{1/2} \qquad (3)$$

where C is a constant, P is the laser power and t is the exposure time.

Similarly the voxel length *l* can be calculated assuming the following axial distribution $N(z) = N_0 / (1 + z^2/z_R^2)$, where $z_R$ is the Rayleigh length :

$$l(P,t) = 2z_R[(CP^2t)^{1/2} - 1]^{1/2} \qquad (4)$$

The resolution of the two-photon polymerization process was investigated by producing individual voxels for different exposure times at a given power. The resin was NOA81. It was found that the threshold power for this resin is 15 mW average power.

In figure 3 is shown a SEM image of a single voxel produced with 15 mW and 60 ms exposure time. It has a diameter of 270 nm and a length of 1.4 μm.

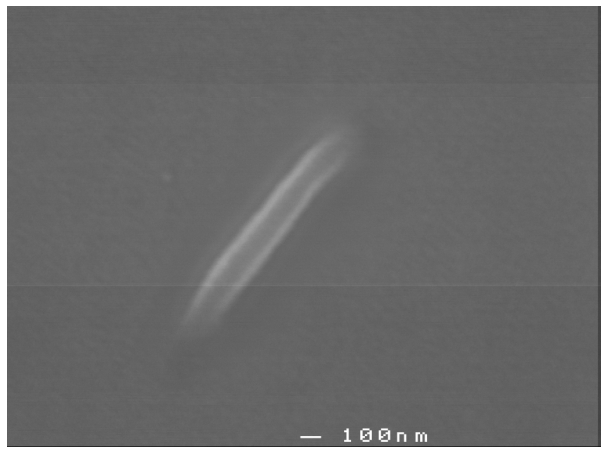

**Figure 3.** SEM image of a single voxel.

At the threshold power, a series of voxels were produced with different exposure times ranging from 40 ms to 500 ms. The smallest voxel has a diameter of 210 nm and a length of 970 nm.

Their diameters were measured from the SEM images and plotted against the exposure time. In order to fit the data we used a function similar to that in equation (3), $f(t) = A_d*[\ln(B_d*t)]^{1/2}$. The program finds the unknown constants A and B corresponding to the best fit. The plot of the fit is shown in figure 4.

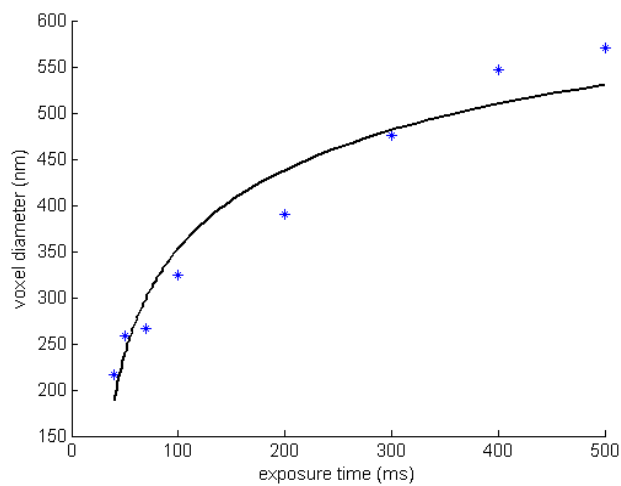

**Figure 4.** Measured and estimated voxel diameters as a function of time for NOA 81 resin

In this case $A_d = r_0 = 312$ nm and $B_d = 0.036$ s$^{-1}$. The estimated parameter $r_0$ is very close to the theoretical value given by the lateral resolution of the objective $r = 0.61\lambda/N.A. = 366$ nm [19].

In the same manner we plotted the voxel lengths against exposure times and fitted the data with a function of the form $f(t) = A_l*[(B_l*t)^{1/2} - 1]^{1/2}$ (figure 5). For the coefficients we obtained $A_l = 2z_R = 1.74$ μm and $B_l = 0.042$ s$^{-1}$. Again we found that the estimated parameter $2z_R$ is close to the theoretical value of the axial resolution $z = 2\lambda n_{oil}/N.A.^2 = 1.4$ μm [19]. Also for a given power the two coefficients $B_d$ and $B_l$ should have the same theoretical values and we find that the estimated coefficients have close values, proving that the theoretical model holds for our experiments.

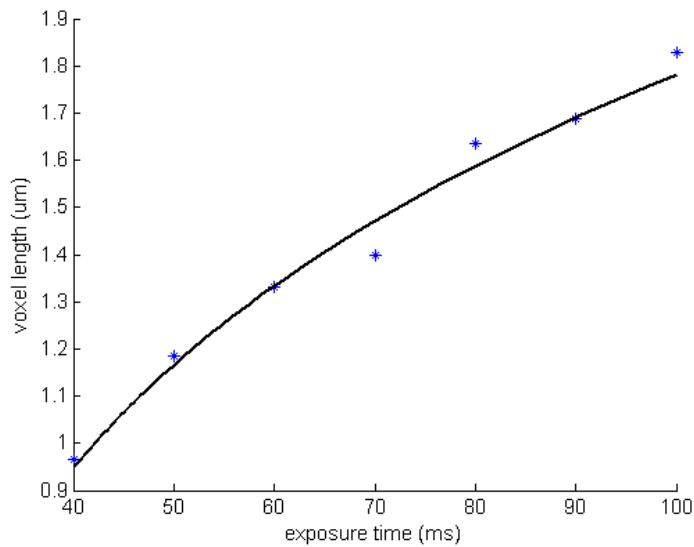

**Figure 5.** Measured and estimated voxel lenghts as a function of time for NOA 81 resin.

Based on this model it is interesting to see how the voxel diameter and length vary as a function of laser power and exposure time (figure 6, $P_{th}$ is the threshold power).

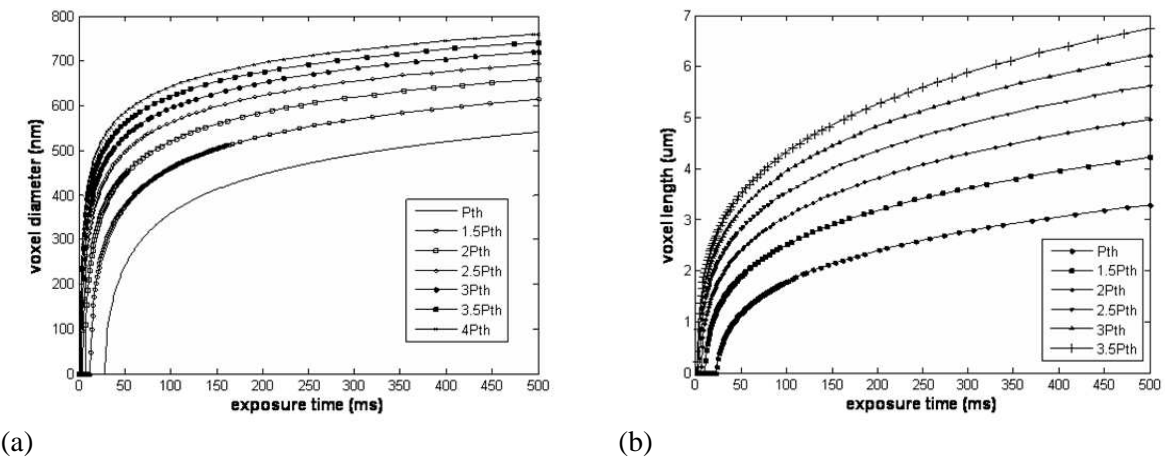

(a)     (b)

**Figure 6.** Variation of voxel diameter (a) and length (b) with the laser power.

It can be clearly seen that in order to achieve high spatial resolution the two-photon polymerization has to take place in the near threshold region and for short exposure times.
Also, the voxel length (axial resolution) is more sensitive to the power than to the exposure time. The voxel diameter (lateral resolution) varies significantly with exposure time up to around 100 ms, above 100 ms being relatively stable.

**4. Microfabricated structures**

We are interested in microscopic objects which can be 3D optically trapped and can transfer optical angular momentum. The micro-structures produced by two-photon photopolymerization are suitable for trapping experiments due to their low absorption of the IR trapping beam and the refractive index of the polymerized resin (n = 1.56).
On the other hand, in order to obtain transfer of angular momentum between the trapping beam and the trapped object, and hence torque, one needs to take into account the shape of the trapped object. As discussed in [20], the most important aspect of the shape of the object is its rotational symmetry. Thus the rotational symmetry of the object can be tailored in order to optimize the torque. So, depending on the angular momentum of the incident beam one can find a suitable order of rotational symmetry for optimum torque efficiency. Furthermore, if one can produce chiral objects, torque can be achieved by using incident beams carrying no angular momentum. At the same time, by using chirality, one can design a microscopic optical element which can add orbital angular momentum to a normal Gaussian laser beam.
Here we show two sorts of devices which we microfabricated and illustrate the above ideas.

*4.1. Off-set rotor with 4-fold rotational symmetry*

We designed and microfabricated a chiral object namely an off-set cross with a stalk that would give the proper alignment in the optical trap. Shown in figure 7 are the CAD model of the objects and the bitmaps which are input into the scanning program.

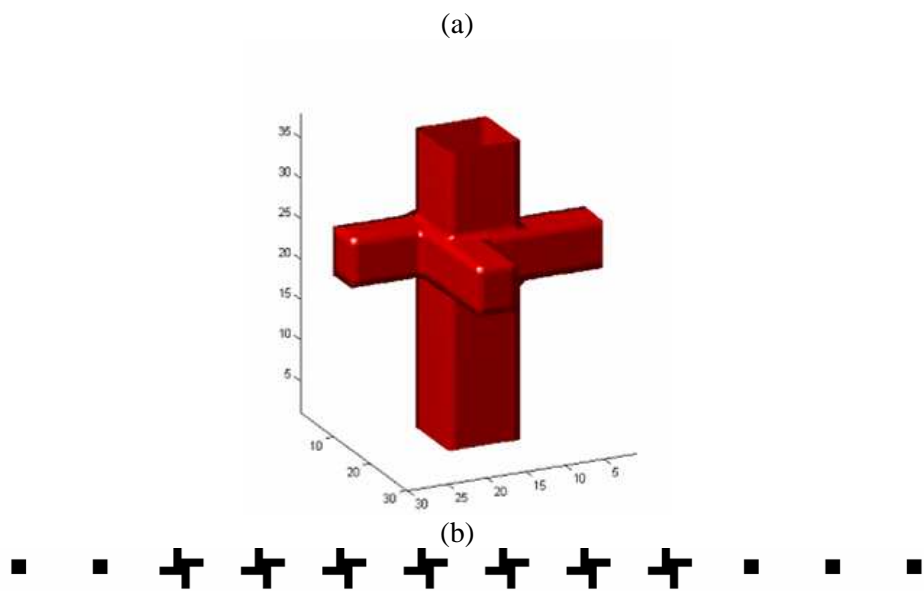

**Figure 7.** 3D design of the microrotor (a) and sequence of bitmaps defining the 3D object (b).

The size of the squares defining the stalk is chosen to be 18X18 pixels and the off-set cross is 54X54 pixels. The whole object is composed of 41 layers giving a physical length of 8 μm. The parameters for successful resin polymerization are 25 mW of infrared laser power entering the microscope, 80 fs for the pulse duration at a repetition rate of 80 MHz and 8 W of pumping laser power at a scanning speed of 15 μm/s.

Optical microscope images of the produced structures in unpolymerized resin and after rinsing with acetone are shown in figure 8.

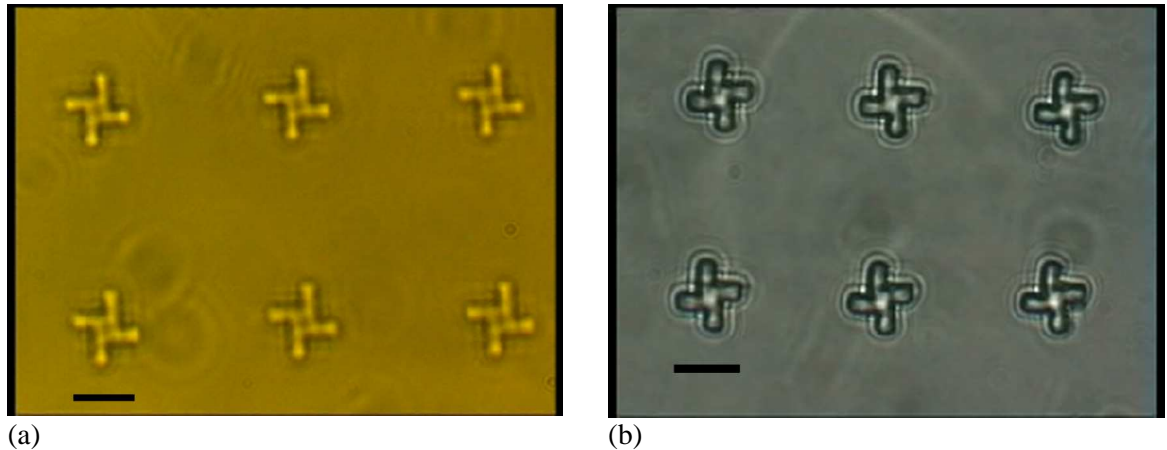

(a)                                                         (b)

**Figure 8.** Optical microscope images of the 3D off-set crosses produced by two-photon polymerization in unpolymerized resin (a) and after rinsing with acetone (b) (scale bars are 5 μm).

Due to the high travel range of the scanning stage one could produce a large number of microstructures in one go which is a big advantage in terms of fabrication efficiency. The fabrication time for each structure is about 15 minutes.

The structures were also imaged with SEM, which is a powerful tool due to its high resolution in the nanometer range. A typical SEM image of the fabricated structure is shown in figure 9.

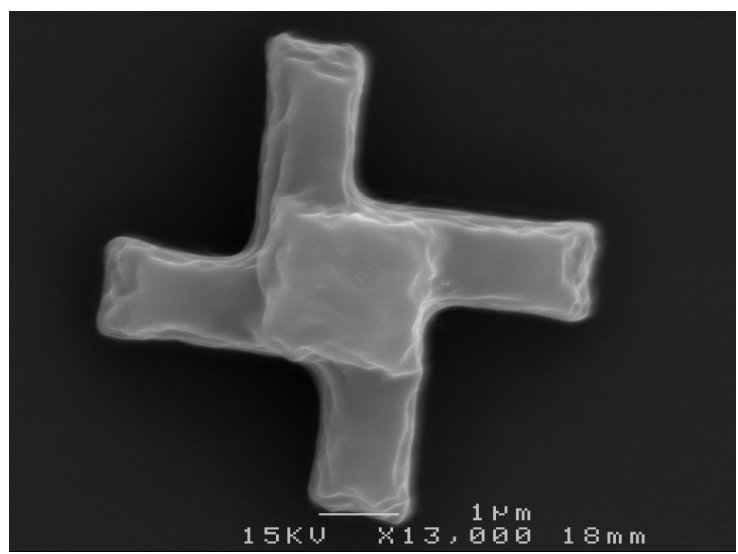

**Figure 9.** SEM image of the 3D off-set cross.

The layer-by-layer fabrication discussed above can be clearly seen from the SEM image as well as the resemblance to the CAD model. It is of interest to note that the shape of the object after acetone washing is dependent on the amount of acetone used for rinsing the unpolymerized resin. In order to obtain a structure with smooth surfaces only three to five drops of acetone need to be used otherwise the structure is corroded by the surplus amount of acetone. These objects are released from the cover slip by pushing them with a needle attached to a translation stage. They were trapped in water using a linearly polarized Gaussian beam and they rotated at a frequency of 10 Hz with 150 mW of laser power. Since the incident beam carried no angular momentum, the rotation was clearly achieved by scattering of the Gaussian beam into $LG_{04}$ modes by the microrotor.

*4.2. Microscopic diffractive optical element (MDOE)*

We have also fabricated microscopic diffractive optical elements. These objects are of special interest to us because they can be designed to create orbital angular momentum for optical trapping beams. Essentially they behave as micro-holograms similar to the computer generated holograms [21]. The advantage of these micro-objects is that they can be mass produced and can be integrated into lab-on-chip type experiments.

We have designed and fabricated an MDOE with 8-fold rotational symmetry. It consists of eight saw teeth with a height calculated such that, when the MDOE is immersed in water, induces a $2\pi$ ramp in the phase front of the incident beam. Hence, due to the 8-fold symmety, $LG_{08}$ modes can be created carrying orbital angular momentum.

In figure 10 are shown the CAD model and an SEM image of the MDOE with a diameter of 9 μm and a height of 5 μm. This object, as schematically shown in figure 10a, is built up in layers as the off-set cross. The actual layers can be clearly seen in the SEM image in figure 10b. The photopolymerization parameters were the same as with the fabrication of the off-set cross.

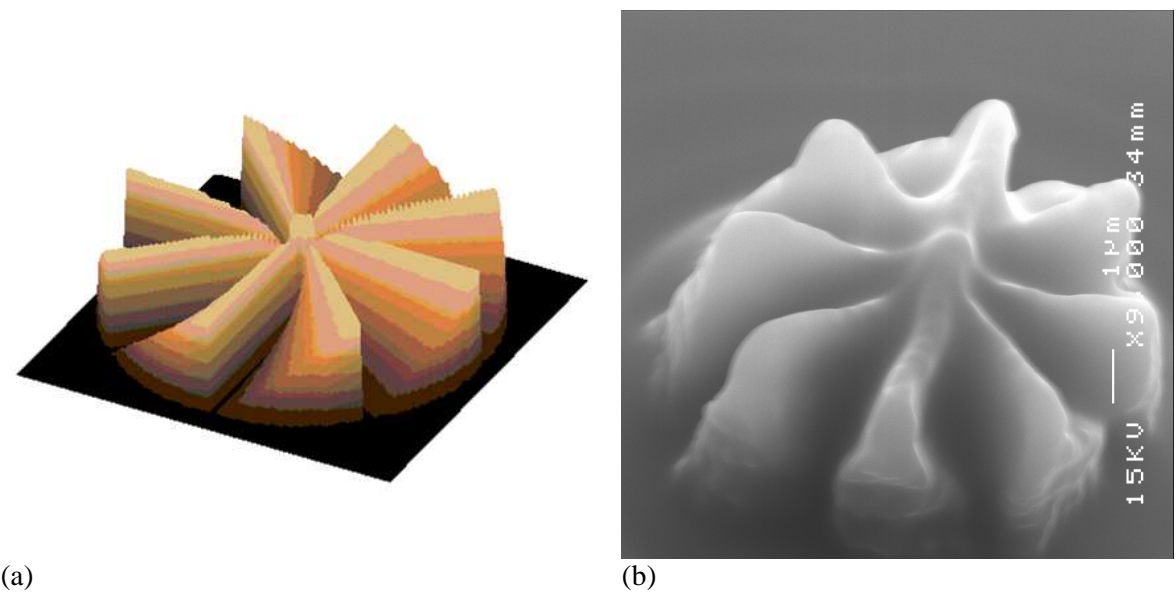

(a)      (b)

**Figure 10.** 3D design of the MDOE (a) and an SEM image of the MDOE (b).

It has been shown [22] that the above described MDOE induces orbital angular momentum to a Gaussian trapping beam thus rotating non-chiral and non-spherical rotationally symmetric micro-objects.

**5. Conclusion**

Two-photon photopolymerization of optically curable resins is a powerful method of fabricating 3D micrometer size structures of arbitrary shape. The high lateral resolution of this process, below the diffraction limited resolution of light, allows fabrication of micro-objects with fine features. Even though the axial resolution of the process is inherently lower than the lateral resolution, it can be reduced by setting a small offset between subsequent layers.
We have shown the applicability of the method by fabricating micro-structures suitable for optical trapping experiments. We have also shown that this technique allows fabrication of devices which can generate orbital angular momentum from Gaussian laser beams being at the same time compatible with lab-on-chip applications.